# Stimulating Entrepreneurship in Teaching Human Computer Interaction

**Dusanka Boskovic[1], Nihad Borovina[2], Merima Zukic[3]**

[1]University of Sarajevo, Faculty of Electrical Engineering
[2]BH Telecom dd
[3] University of Sarajevo, Faculty of Philosophy
E-mail: dboskovic@etf.unsa.ba

**Abstract.** *Software development requires understanding of users, user needs, user tasks and context in which they are operating. These skills are familiar to entrepreneurs, product managers, and marketing experts. However, our teaching experience suggests that students generally find these topics less attractive as they perceive them to be far too theoretical and thus, not as useful. During the years of teaching the Human Computer Interaction course we have noticed students' preferences for learning technology oriented methods, or what we refer to topics belonging to solution domain. The changes in the modernized HCI course introduced Product Market Fit canvas in order to bridge the gap between "theoretical" and "practical" part of the course.*

## 1   Introduction

The Human Computer Interaction (HCI) course is focused on designing software/system interaction as an approach to develop usable software/system. The HCI course is taught at the master cycle as an elective for students studying Control and Electronics, Informatics and Telecommunications. Students gain knowledge on designing interaction for complex systems such as: SCADA systems, systems for safety and signalization, monitoring and supervision of networks, etc. The process of designing user interaction places a user in the research focus, and user research draws upon good practices from the product development [1, 2]. Interesting example of documenting and managing product requirements based on the user needs is Product Market Fit (PMF) canvas [2]. It enables a developer to directly link problem domain concepts to solution domain.

The Erasmus+ KA2 project BENEFIT (2017-21) [3] has been aimed at improving engineering education in the area of ICT and our course was included in the modernization. In the academic year 2018/2019 we have introduced the PMF canvas, with adjustments for developing software product/service [4]. With this improvement we wanted to stimulate a range of entrepreneurial skills that successful interaction designer requires: critical thinking, multidisciplinary orientation, creative problem solving.

The rest of this paper is organized as follows. Section 2 presents a brief overview of competences related to the entrepreneurship, and appropriate delivery and assessment methods. Section 3 is addressing students' preferences towards learning technology oriented methods, the difference in students' attitude towards topics belonging to solution domain opposite to topics belonging to problem domain. Section 4 describes Product Market Fit canvas, a tool used for documenting and managing product requirements based on user needs. Section 5 briefly describes the experience and knowledge gained from the Teachers Training workshop organized within the framework of the BENFIT project. Section 6 summarizes the discussion and conclusions.

## 2   What didactics teaches us?

Bellotti et al. state that entrepreneurship is a personal skill and motivation which draws a person to engage his/her abilities and efforts in the creation of a new product and services, with a business value given by the ability to match the market demand [5].

We can use the previous statement to interpret entrepreneurship as a generic skill, since it is not connected to specific field of study. Also, we can think about entrepreneurship as a transferable skill that can be used in all of domains of human functioning. Having said this, market demand and business value of activity depend on the specific context in which person exists. It is not necessary to generate large income or own a company for one to be considered a good entrepreneur.

In regards to entrepreneurship education, it was said that it could not be performed in HEI optimal way, since planning is incompatible with entrepreneur - friendly society [6]. However, as the lifelong learning and personal development are becoming increasingly more relevant and important issues, the entrepreneurship education is gaining popularity, not only with HEI, but also within the society, in general. It is becoming increasingly clear that entrepreneurship should be integrated through the entire education system.

There are a variety of difficulties when it comes to integrating entrepreneurship into standard curricula. It is a transferable and an interdisciplinary skill that needs to take into account pedagogical and psychological knowledge and strategies. At most universities, a primary focus is placed on vocational competency, while entrepreneurial skills are seen to be of secondary importance. There is an issue regarding competences of academic staff for teaching problem solving and critical thinking, and these skills are essential for entrepreneurship. There is sometimes a conceptual ambiguity both on how to teach and how to evaluate development of entrepreneurial skills.

Entrepreneurial approach supports a range of different skills and abilities that are related to creativity, problem solving, positive "to do" attitude, bravery, innovation. It has a multi–disciplinary orientation.

Development of entrepreneurial skills needs to be defined in learning outcomes, reflected on teaching techniques and planned in evaluation. In didactics we call this constructive alignment,

There are different teaching strategies that can be used to foster entrepreneurial skills, but what is important is to plan lessons with this goal in mind and try to engage students in problem solving and critical thinking, no matter the specific topic and strategy.

Assessment tools need to be in-line with learning goals and teaching techniques. Sometimes, classic assessment strategies are not appropriate to measure the level of entrepreneurial skills acquisition.

|  | Excellent | Competent | Need improvements |
|---|---|---|---|
| Task description |  |  |  |
| Dimensions |  |  |  |
| Scales |  |  |  |
| Description of each level |  |  |  |

Figure 1. Rubric example

Thus, a special attention needs to be devoted to assessment techniques. For this purpose, we have selected Rubrics [7]. This is a special scoring tool that lays out specific expectations for the assignment. The Rubrics are very useful when we want students to develop the ability to appropriately reflect on ill-structured problems. Rubrics divide an assignment into a number of distinct parts and provide a detailed description of what constitutes an acceptable or unacceptable level of performance for each of those parts.

Rubrics include a task description (the assignment), a scale of some sort (levels of achievement, possibly in the form of grades), the dimensions of the assignment (a breakdown of the skills/knowledge involved in the assignment), and descriptions of what constitutes each level of performance (specific feedback) all set out on a grid. In fig. 1, an example of a rubric used during the BENEFIT project teachers training is presented. Participants were asked to develop a rubric as a grading tool for an assignment from their course, and the assignment expectations were defined as a rubric.

## 3 What we have learned from experience?

During the years of teaching the Human Computer Interaction course we have conducted student evaluation surveys to use data for continuous improvement. The survey is addressing the importance of topics covered in the course and students reflection on their achieved results. We have notices students' preferences for learning technology oriented methods, and topics belonging to solution domain. The results suggest that students find topics belonging to problem domain more theoretical and less attractive and this is related to their perception of usefulness of these topics.

HCI syllabus includes the following topics: usability principles, methods and tools for designing interaction and user experience, prototyping techniques, user interaction design patterns, and evaluation of usability. Traditional lectures covering theoretical principles are followed by the lab exercises, homework assignments and a final project. There is no written nor oral exam. Students are engaged in a project to design and implement an interactive application by using software development technology of their own choice.

Our objective was to find an appropriate learning activity that is able to bridge the gap between "theoretical" and "practical" part of the course. Developing a successful product requires understanding of user needs, user tasks, as well as the context in which they are operating. These skills are familiar to product managers, business analysts and marketing experts. Product Market Fit (PMF) canvas is an interesting example of documenting and managing product requirements based on user needs. It directly links problem domain concepts to solution domain. In this academic year we have introduced this tool, with adjustments for developing software product/service.

After the course completion, we organized the survey again, with two questions for each topic, assessing how useful and how fun it was for students (5 point Liker scale). The topics assessed: Cognitive concepts and interaction design principles, User and user needs analysis methods, Task analysis methods (HTA), UX –Product Market Fit canvas, Prototyping – paper and mockup tools, Interaction design patterns, Visualization, Usability and user experience evaluations. In the end of the survey we asked students to pick up the three concepts they have found the most important for Interaction Design and also to evaluate the level of their competences.

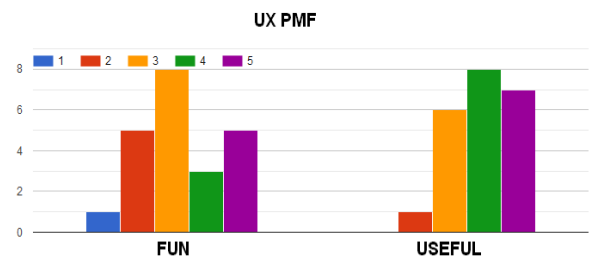

Figure 2. Evaluation of UX PMF Fun vs Useful: Mean values (3.45) vs (4.09)

The analysis of the survey results was conducted. It shows that students have generally found problem domain topics interesting, but less useful. The results for solution domain topics are opposite. The least fun topics were: UX –Product Market Fit canvas (3.45), HTA

(3.81), Prototyping (3.91) and Interaction design patterns (3.91), but Design patterns were graded as the most important (4.55), with Mockup tools following (4.45). Importance of Cognitive concepts and interaction design principles is graded (4.09) and UX – Product Market Fit canvas (4.09).

When selecting the most important topics 80% of students selected Interaction Design Patterns, followed by Error prevention (60%) and Iterative development (50%), User involvement (40%), Responsive design (40%) and Minimalism (20%). Interesting result is that no one selected Visual design and User help.

In assessing the level of their competences, we have offered a scale corresponding to simplified version of the revised Bloom taxonomy [8]. Students evaluated their competences after the course completion on the higher levels: at the level of application (half of the students) and at the level of evaluation (other half).

## 4 Product-market Fit (PMF) concept adapted for teaching HCI

Interaction design is part of product development and teaching HCI has a lot in common with teaching product development, especially in the world of digital products, smartphone applications and web portals. Efficient and simple interaction make these product successful on the market.

One of the most important goals of product development is to achieve product-market fit, meaning being in a good market with a product that can satisfy that market. Product Market Fit pyramid [9] is actionable framework for product development which decompose the process into five components: (1) define target customers, (2) identify underserved needs, (3) build value proposition, (4) specify feature set, (5) do UX design.

Product development is a team activity, usually with many participants. One of the challenges in such process is to ensure that all participants understand the objectives and the business model in the same way. Nowadays, characterized by fast market changes, a large number of innovations, demanding users and high competition, visual tools are increasingly used in order to represent the key aspects of product development process in a simple, concise and clear manner.

Product Market Fit canvas is a tool that supports product-market fit pyramid. Around 150 Lean product development specialists participated in the development of this tool [9]. The most important parts of the Business Model Canvas [2] are the relationship between the Value Proposition (what we are building) and the Customer Segment (what users need). These two components of the business model are so important that can be isolated as a separate canvas with its own name: "Product/Market Fit." [10]

The left side of the PMF canvas is dedicated to the problem domain, while the right side is designated to the solution domain. Each field in the right column is related to a field in the same row in the left column. The version adapted for describing the software application is presented in Fig. 3. Customer is User, Product is Application, and we discard the distribution channel.

The central topic of our HCI course is a fully functional application development, where a particular emphasis is placed on interaction design. Students work in teams and work through all steps of product development. The first step to propose an idea that is realistically applicable in the market. After preliminary approval, the idea is subsequently presented in more details and the final project topic is negotiated. During the development, students receive feedback during regular consultations. Final detailed evaluation and grading is also referring to the PMF.

In that sense, Product Market Fit canvas is intended:
1. To prevent misunderstandings in defining scope and goals of assignments
2. To improve control of the process of application development
3. To emphasize the importance of dealing with the problem space
4. To focus on WHY and WHAT in application development

This approach provides a sense of realism. However, it is also has some limitations. There is not much room for introducing a new topic in the tight course schedule, so the PMF is covered within one week. That is why it is important to have a simple and usable tool for a quick introduction to the topic.

If more time could be allocated to the problem domain, further research of user needs would be conducted using field surveys, contextual inquiries, etc.

The primary goal of using Product Market Fit canvas was to engage students with the problem. However, in practice we have noticed the benefit of its application in several phases of students' work. The main guidelines for Product Market Fit canvas use within the HCI course are:

### A. Left-hand side of canvas is used to submit the project proposal

Students are proposing the topic of their project at the beginning of the semester. The fields on the left side of the PMF canvas describe users, their needs, requirements, and expected outcomes, which is quite enough to encourage students to analyze how much their idea may fit the market needs.

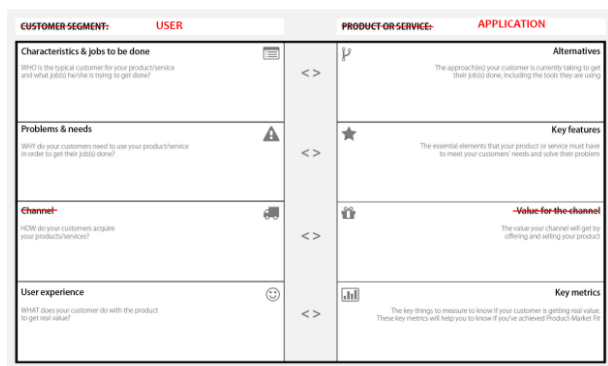

Figure 3. PMF template adapted for describing students' project

*B. Full canvas is used for discussion and project approval process*

This is followed by an approval process which is activity of detailed presentation of the full canvas to professor, assistants and colleagues a few weeks later. Each field of the canvas is presented in detail. Due the simplicity and the clarity of this form, this type of presentation can foster a very productive and meaningful discussion. This activity is already included in the grading scheme.

*C. Canvas is used during the consultancies and for the final project evaluation*

Once the topic is approved, the canvas is archived and later used as a reminder of the objectives during the student consultation hours. Finally, canvas is also used in the evaluation of the final project and to assessing how well the final results have met the goals set at the beginning.

Documenting achieved level of engagement is not an easy task, but level of project completion was raised after introduction of this approach. Percentage of students completing the project during the first exam term raised to 44% (2018/2019) from 37% (2017/2018) and 38% (2016/2017).

## 5 Teachers training experience

As a part of the BENEFIT project [3], teachers training "Designing a Course for Stimulating Entrepreneurship in Higher Education" was organized by the University of Sarajevo, where we shared our experiences with the project partners. It was a two-day training with the first day dedicated to didactics and the second day to teaching product development using the PMF. There were 50 participants on each day. Event evaluations were positive with assessment grades in range 4.3 – 4.8.

Traditional education does not correlate well with the development of entrepreneurial skills, and the topic is rarely adequately addressed in the engineering study programs. Entrepreneurship is quite often considered only in the narrow, commercial form, but it could also be social, public and corporate. The latest form is focused on internal innovation, endeavor and strategic renewal [11] what is all important for a modern engineer.

## 6 Conclusions

The results presented indicate that stimulating the entrepreneurial skills is an appropriate approach for the HCI course. Product Market Fit canvas is a tool we used to help students bridge the gap between "WHY" and "HOW," between the problem domain and the solution domain. The survey results provided the evidence that students recognized importance and benefits of this tool. The improvement in number of timely completion and submission of the students' projects is backing up this conclusion.

This topic was included in the teachers training within the framework of the BENEFIT project, with relevant didactic topics included. The participants' feedback was positive and indicating broader interest in introducing activities that will stimulate entrepreneurial attitude in engineering students.

## Acknowledgements

"This work was carried out within the project BENEFIT - Boosting the telecommunications engineer profile to meet modern society and industry needs - 2017-2021 - cofounded by the ERASMUS+ KA2 program – grant agreement n. 585716-EPP-1-2017-1-AT-EPPKA2-CBHE-JP".